\documentclass[aps,groupedaddress,superscriptaddress,amsmath,amssymb,prb]{revtex4-2}

\usepackage{bm}
\usepackage[pdftex]{graphicx}
\usepackage{xcolor}
\usepackage{color}
\usepackage[normalem]{ulem}
\usepackage{enumerate}
\usepackage{epstopdf}
\usepackage{braket}
\usepackage{hyperref}
\usepackage{blindtext}
\usepackage{comment}
\usepackage{amsmath}

\begin{document}
%\linenumbers

\title{Spin dependent fluorescence mediated by anti-symmetric exchange in triplet exciton pairs}

\author{Yan Sun}
\affiliation{LPS, Universit\'e Paris-Saclay, CNRS, UMR 8502, F-91405 Orsay, France}
\author{M. Monteverde}
\affiliation{LPS, Universit\'e Paris-Saclay, CNRS, UMR 8502, F-91405 Orsay, France}
\author{V. Derkach}
\affiliation{O. Ya. Usikov Institute for Radiophysics and Electronics of NAS of Ukraine 12, Acad. Proskury st., Kharkov, 61085, Ukraine }
\affiliation{LPS, Universit\'e Paris-Saclay, CNRS, UMR 8502, F-91405 Orsay, France}
\author{T. Chaneli\`ere}
\affiliation{Univ. Grenoble Alpes, CNRS, Grenoble INP, Institut N\'eel, 38000, Grenoble, France}
\author{E. Aldridge}
\affiliation{Department of Chemistry, University of Kentucky, Lexington, KY 40506-0055, USA }
\author{J. E. Anthony}
\affiliation{Department of Chemistry, University of Kentucky, Lexington, KY 40506-0055, USA }
\author{A.D. Chepelianskii}
\affiliation{LPS, Universit\'e Paris-Saclay, CNRS, UMR 8502, F-91405 Orsay, France}

\begin{abstract}
Singlet fission and triplet-triplet annihilation (TTA) are spin-dependent phenomena critical to optoelectronics. The dynamics of spin populations during geminate triplet pair separation are crucial for controlling fission and TTA rates. We show that the Dzyaloshinskii-Moriya interaction (DMI) induces level crossings between spin manifolds, affecting spin populations and TTA rates in crystalline fission semiconductors. By investigating spin-dependent fluorescence in a triplet exciton pair with the magnetic field aligned along the fine structure tensor, we isolate the effect of DMI, as the triplet wavefunctions remain unaffected by the field. Our results reveal that DMI introduces additional TTA pathways that are forbidden by spin conservation, explaining the observed evolution of optically detected magnetic resonance signals with varying magnetic field. This study highlights the significant impact of DMI on the optical properties of triplet excitons, advancing our understanding of spin dynamics in these systems.
\end{abstract}

\maketitle

Spin-dependent optical processes are ubiquitous in organic materials with various manifestations depending on the physical context. For electron-hole pairs, the mechanisms differ between bound excitons and free carriers. In bound excitons, intersystem crossing between triplet and singlet manifolds plays a crucial role in enhancing the efficiency of organic light-emitting devices \cite{tang1987organic,burroughes1990light,friend1999electroluminescence,baldo2000high,uoyama2012highly,zhang2012design, matsushima2019high, song2020organic, li2022singlet}. It is essential for the detection of magnetic resonance signals from a single spin  \cite{schmidt1968optical,clarke1982triplet,wrachtrup1993optical,kohler1993magnetic,shinar2012optically} and the development of room-temperature microwave amplification by stimulated emission of radiation (MASERs) \cite{oxborrow2012room,bogatko2016molecular}, as well as the applications in magnetometry \cite{mena2024room}. For separated electron-hole pairs, spin mixing is responsible for the high field sensitivity of organic magnetoresistance devices \cite{mermer2005large,sheng2006hyperfine,bobbert2007bipolaron,wang2012control,harmon2014organic,gobbi2017rise} and may offer insights into magnetoreception in bird navigation \cite{wiltschko2005magnetic,rodgers2009chemical,mouritsen2018long}.
Pairs of triplet excitons also exhibit rich spin-dependent physics, such as TTA and singlet fission. Singlet fission has garnered attention for its potential in photovoltaics \cite{Hanna2006,tayebjee2012,Ehrler2012,lee2013,Congreve2013,Dirk2018,Einzinger2019}, light emitting diodes \cite{di2017efficient} and upconversion \cite{Keivanidis2003,Singh2010,Schmidt2010,Pandey2015,Wenping2021,Bossanyi2021}. In this process, a singlet exciton splits into two lower-energy triplet excitons, leading to the formation of transient bi-excitons with higher spin states (S=2), which serve as intermediates between the photoexcited singlet exciton and the resulting pair of triplet exciton \cite{Tayebjee2017,Weiss2017,bayliss2018,bayliss2016spin,wakasa2015can,chakraborty2014massive,huang2021competition,yago2022triplet}.

The difference in optical lifetimes of the spin eigenstates allows precise identification of the microscopic structures of triplet pairs \cite{ODMR2019,sun2024cascade}. However, the mechanisms underlying the role of spin in geminate triplet pair separation remain incompletely understood. Current models suggest a combination of spin-mixing dynamics, analogous to radical pairs where only the singlet component is optically active, and mechanisms similar to intersystem crossing for triplets, where spin-orbit interaction plays a significant role \cite{sun2023spin,kraffert2014charge,budil1991chlorophyll,el1972pmdr,Singh1965}. The fluctuations in the exchange interactions were shown to increase the rate of quintet formation\cite{collins2023quintet}. 
In this work, we demonstrate how the antisymmetric exchange (Dzyaloshinskii-Moriya interaction, DMI) \cite{dzyaloshinsky1958thermodynamic,moriya1960anisotropic} becomes relevant for pairs of triplet excitons by inducing level anticrossings between singlet and triplet manifolds of the triplet pair that occur when a geminate triplet pair separates. We show that this interaction explains the evolution of the optically detected magnetic resonance (ODMR) signal as a function of the magnetic field. In our experiments we study crystals of bis(triisopropylsilylethynyl)anthradithiophene (TIPS-ADT) \cite{dean2017photophysical,sundaram2024polymer} for which the magnetic field is aligned along the fine-structure anisotropy axis. This avoids changes in the single-spin eigenfunctions with varying magnetic fields.

Crystals of TIPS-ADT contain only one molecule per unit cell, enabling a detailed investigation of the microscopic mechanisms responsible for the ODMR contrast. For this purpose, we deposited a TIPS-ADT crystal on a microwave stripline. An optical fiber positioned directly above the crystal provides the optical excitation and fluorescence collection \cite{sun2024cascade}. To align the applied magnetic field with the molecular axes of the system, we utilized a two-axis superconducting magnet in combination with a goniometer. The goniometer allows for the rotation of the sample along the vertical axis, thereby aligning the main fine structure tensor direction within the plane of the vector magnetic field. Through the compensation of the in-plane field, this configuration permits the ramping of the magnetic field along a specified direction,which in the context of this study corresponds to the direction of the fine structure tensor. The sample and the stripline are cooled to a temperature of 4K thermalized to the helium bath using Helium gas vapor.

\begin{figure}
\includegraphics[clip=true,width=0.7\columnwidth]{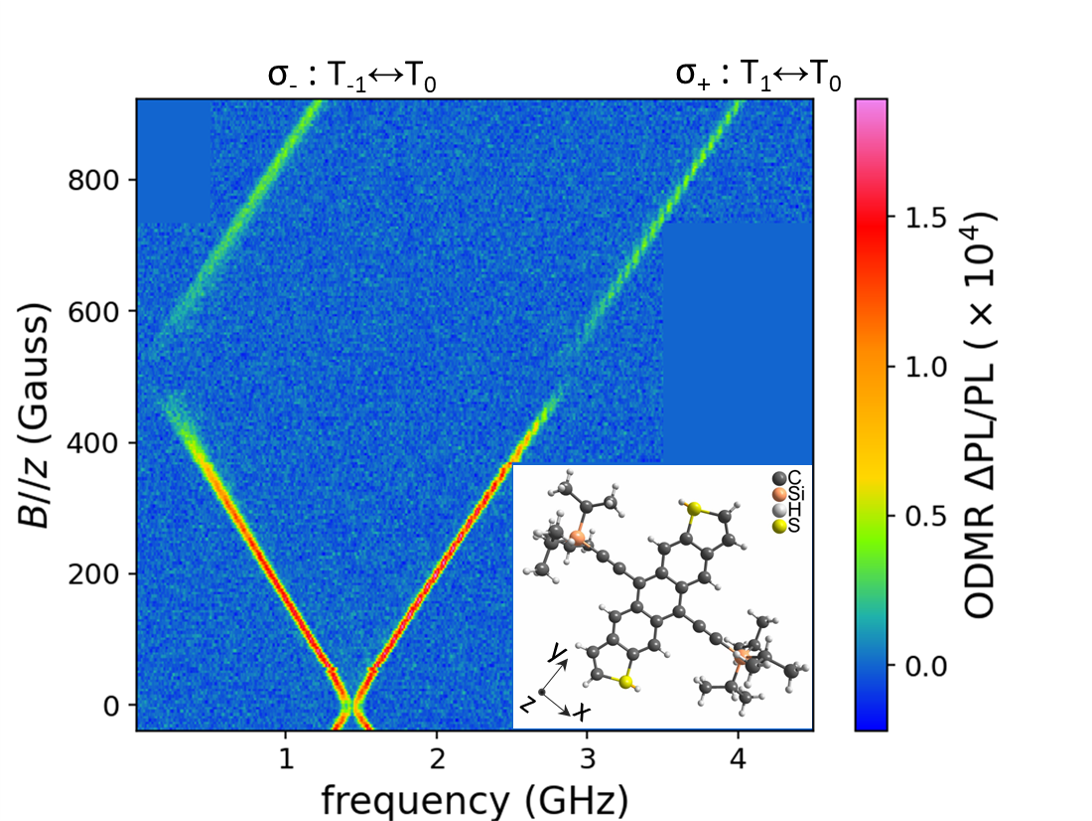}    
\caption{Optically detected magnetic resonance (ODMR) signal from a TIPS-ADT crystal was measured with the magnetic field aligned along the $z$ axis. The inset displays the molecular structure of TIPS-ADT. Two transitions, denoted $\sigma_{\pm 1}$ are observed between the $|T_0\rangle$ and $|T_{\pm 1}\rangle$ triplet spin states, which are split by the Zeeman effect in the presence of a magnetic field. The ODMR amplitudes for the $\sigma_{\pm 1}$ are approximately equal across all the magnetic fields. Both amplitudes vanish at the level crossing between $|T_0\rangle$ and $|T_{-1}\rangle$ at $B = D$.  Additionally, a reduced ODMR amplitude is observed for $B > D$ in comparison to $B < D$. }
\label{odmrmap}
\end{figure}

In Fig.~\ref{odmrmap} we show a map of the ODMR signal as functions of the magnetic field applied along the molecular $z$ direction (see inset Fig.~\ref{odmrmap} ) and microwave frequency. This figure displays a textbook picture of the Zeeman effect splitting the $|T_{\pm1}\rangle$ triplet exciton when the magnetic field is applied along the main zero-field splitting anisotropy axis. We label $|T_{0, \pm 1}\rangle$ the three spin states of an isolated triplet. This leads to two spectroscopic lines $\sigma_{\pm 1}$ corresponding to transitions between $|T_0\rangle$ and $|T_{\pm 1}\rangle$. The ODMR signal comes from the different optical contrast between $|T_{\pm1}\rangle$ and $|T_0\rangle$ spin eigenstates similar to what can be seen with color centers in solids \cite{awschalom2018quantum,barry2020sensitivity,bhaskar2020experimental}. However, the precise dependence of the ODMR signal amplitude on the magnetic field is rather subtle. As the magnetic field increases the $|T_0\rangle$ and $|T_{-1}\rangle$ states cross at $B \simeq 500\;{\rm G}$. This corresponds to the magnetic field for which $g \mu_B B = h D$, or $B = D$ if we use the same units for $B$ and $D$ which will be our choice in the following. For $B < D$, the ODMR amplitude seems weakly dependent on the magnetic field; However, it starts to drop as $B$ approaches $D$ and vanishes at the level crossing. For $B > D$ the amplitude increases again but only to about $20\%$ of its value for $B < D$. The drop in ODMR  amplitude is visible when comparing at the same frequency with magnetic fields above and below the level crossing. This is easily observed in the $0.5 - 1$ GHz frequency range, thus ruling out a possible change of the exciting microwave field with frequency. The transition linewidth changes only weakly on both sides of the level crossing. Following the transition $\sigma_+$, we find an average linewidth of around $45\;{\rm MHz}$ for $B < 300\;{\rm G}$ and of  $55\;{\rm MHz}$ for $B > 700\;{\rm G}$. It is then unlikely that the drop in ODMR amplitude is directly related to a change in $\tau_2$ spin transverse relaxation time with magnetic field.

The variation of the ODMR amplitude with magnetic field can have several physical origins. An important mechanism is the change of spin eigenfunctions due to the competition between the Zeeman effect and fine structure or dipole-dipole interactions. In the case of a triplet exciton, the zero-magnetic field eigenstates have different fluorescence yields, and ODMR amplitudes change with the magnetic field due to the mixing among zero-field eigenstates. For a pair of triplet excitons, the pair can recombine radiatively only when the spins of the two triplet excitons are antialigned forming an entangled singlet wavefunction that allows recombination into a singlet ground state without violation of the spin-conservation rule. In both cases, the ODMR amplitude is only determined by the spin eigenfunctions. In our experiment, the magnetic field is aligned along the main anisotropy axis, and the spin eigenfunctions are identical on both sides of the level crossing. This suggests that the ordering of the spin energy levels needs to be reconsidered. If the spin population follows a thermal distribution, the $4.2\;{\rm K}$ Helium bath temperature being much larger than $D$, where the $\sigma_+$ transition would be relatively unaffected, while $\sigma_-$ would vanish at $B = D$. However, such a behavior is not consistent with Fig.~\ref{odmrmap}. 

\begin{figure}
\includegraphics[clip=true,width=0.9\columnwidth]{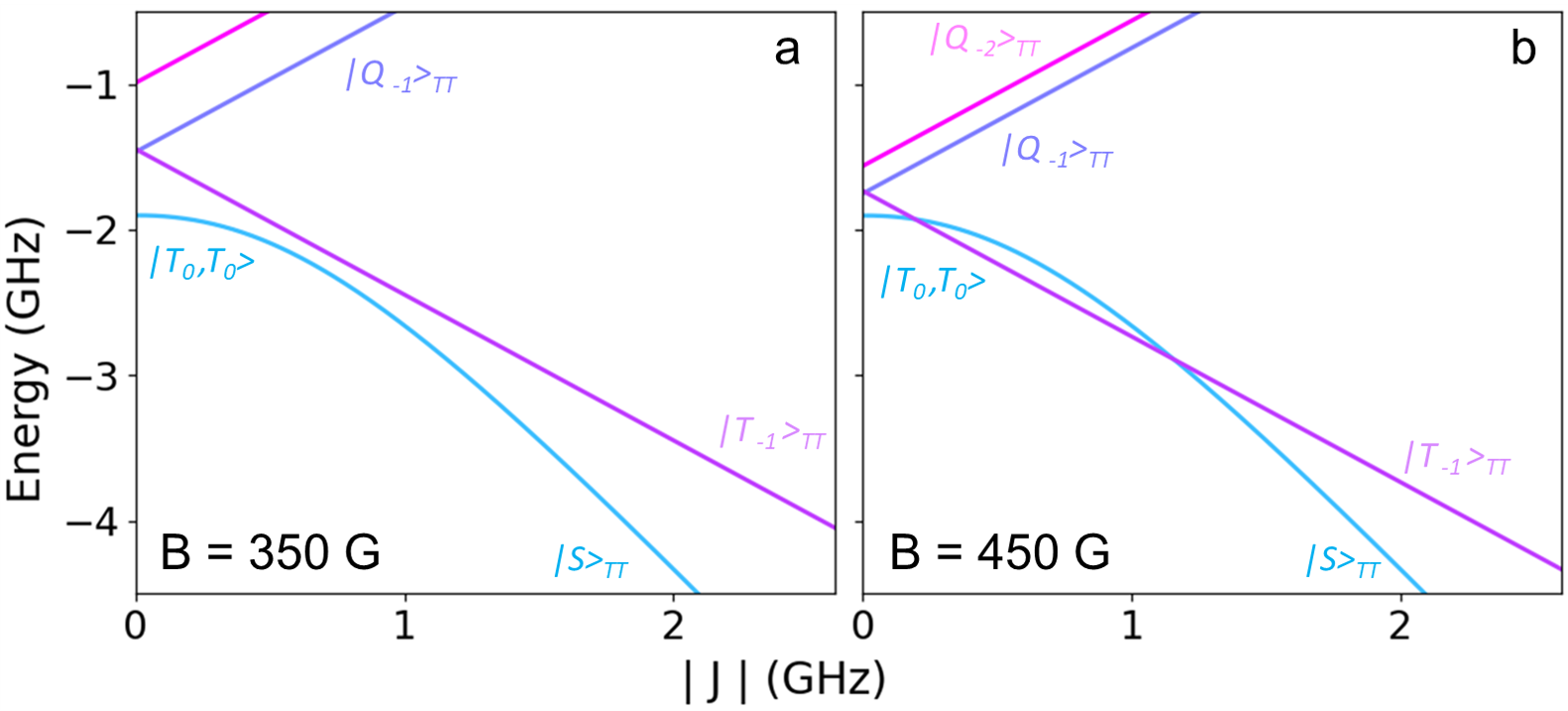}    
\caption{The energy diagram of the lowest energy states of a triplet pair as a function of the exchange energy  $| J |$  is presented for low magnetic field a) $B = 350\ G < 7/9\ D $ and b) $B = 450\ G \approx 8/9 \ D $. For large exchange energy, the states have well-defined total spin configurations, forming singlet $|S\rangle_{TT}$, triplet $|T_n\rangle_{TT}$ and quintet $|Q_n\rangle_{TT}$, where the index $n$ denotes the spin projection. As the exchange energy decreases,  the singlet state adiabatically connects to the separated triplet state $|T_0, T_0\rangle$, resulting in an excess population of the $|T_0\rangle$ state after separation. For $B < 7/9 \ D$, there is no level crossing between the $|S\rangle_{TT}$ and $|T\rangle_{TT}$ states. For $B > 7/9 \ D $, the cross between the $|S\rangle_{TT}$ and $|T\rangle_{TT}$ branches occurs, corresponding to the amplitude of  ODMR signals start dropping.
}
\label{figcrossinglowB}
\end{figure}

Since we are dealing with photogenerated triplets, the population of the triplet spin states can follow an adiabatic population transfer process instead of a thermal distribution. With this mechanism, a triplet pair with a large exchange $J(t)$ is photogenerated in its singlet state $|S\rangle_{TT}$, the subscript emphasizes the triplet pair states (the single triplet exciton states are labeled $|T_{0,\pm1}\rangle$ without subscript). The exchange energy $J(t)$ then decreases as the triplets separate leading to a non-equilibrium spin population which depends on all the level crossings encountered as $J(t)$ goes to zero. This adiabatic population evolution will be sensitive to the level crossing between $|T_{-1}\rangle$ and $|T_{0}\rangle$. However, the drop in ODMR amplitude starts already at $B > 400\;{\rm G}$ relatively far from $B = D$ ($B = 500\;{\rm G}$) when compared to the linewidth of the triplet lines around $20\;{\rm G}$.
To understand the possible origin of this early drop, we plot in Fig.~\ref{figcrossinglowB}.a the energy level diagram of a pair of triplet excitons with identical fine structure as a function of the exchange energy $J$ for $B = 350\;{\rm G}$ along the main anisotropy axis ($D = 1.4\;{\rm GHz}$). This energy spectrum is obtained from the following triplet pair Hamiltonian:
\begin{align}
  \hat{H}_0(t) &= D \sum_{\nu=a,b} S^2_{\nu,z} + B \sum_{\nu=a,b} S_{\nu, z} + {\hat H}_{dip} -J(t) \mathbf{{\hat S}}_{a}\cdot\mathbf{{\hat S}}_{b}
  \label{eq:H0}
\end{align}
It contains respectively the fine structure D terms, the Zeeman B term, dipole-dipole interactions, and a time-dependent exchange energy $J(t)$. The latter describes the dissociation of a triplet exciton pair generated through singlet fission at time $t=0$ in its singlet state $|S\rangle_{TT}$. At $t=0$ when the two geminate triplets are photogenerated, they are strongly coupled, and we assume that the initial exchange energy is antiferromagnetic nature with $|J(t=0)| \gg D$. The exchange energy then progressively drops to zero as the pairs dissociate. For simplicity, we assumed that the rate $dJ/dt$ is constant but fluctuates in a broad range of values depending on the microscopic dissociation process.

At a magnetic field of $B = 350\;{\rm G}$ in Fig.~\ref{figcrossinglowB}a, the singlet pair state $|S\rangle_{TT}$ which is the ground states at large antiferomagnetic $|J|$ adiabatically transforms into the triplet pair state $|T_0,T_0\rangle$ leading to an excess population of the triplet projection $|T_0\rangle$. The excess population in $|T_0\rangle$ with equal $|T_{\pm 1}\rangle$ population can explain equal ODMR amplitudes for both  $\sigma_{\pm 1}$ transitions . While  $|S\rangle_{TT}$ does not cross any other level as $|J(t)|$ decreases to zero, it comes very close to the S=1 triplet pair state $|T_{-1}\rangle_{TT}$. The  $|S\rangle_{TT}$ and $|T\rangle_{TT}$ branches cross at higher magnetic field $B > 7 D/9 \simeq 390\;{\rm G}$ (Fig.~\ref{figcrossinglowB}b). The limit $B = 7D/9$ is very close to the field where the amplitudes in Fig.~\ref{odmrmap} start to drop. This observation suggests that the level crossing between  $|S\rangle_{TT}$ and $|T\rangle_{TT}$ is responsible for the evolution of the ODMR signal with magnetic field.

If selection rules allow mixing with the triplet manifold at this level crossing, the exciton pair is likely to undergo rapid non-radiative TTA. In this scenario, the formation of an excited triplet exciton $T^*$ becomes spin allowed, leading to the creation of an isolated triplet state, which subsequently relaxes to the triplet ground state. This isolated triplet exciton must await until it encounters another triplet to form a non-geminate triplet pair before it can re-emit at its fluorescence wavelength. This process is very slow at 4K due to the low diffusion rates of thermalized triplet excitons. Consequently, the mixing with the triplet pair manifold and TTA reduces the ODMR amplitudes observed in the experiment. In the subsequent discussion, TTA will consistently refer to this non-radiative recombination pathway. However, Fig.~\ref{figcrossinglowB}.b  shows that there is no gap opening at the intersections with $|T_{-1}\rangle_{TT}$ for the Hamiltonian Eq.~(\ref{eq:H0}). Indeed, TIPS-ADT has only one molecule per unit cell, resulting in identical fine structure tensors for both triplets, and the mixing between symmetric (singlet/quintet) and anti-symmetric (triplet) states under the exchange of the two triplets is forbidden. Therefore, additional terms to the spin Hamiltonian are thus needed to open a gap at the level crossings with the triplet manifold.

We show that introducing a finite DMI within the photogenerated pair can explain the observed ODMR amplitudes. The DMI interaction is known to be crucial in stabilizing topological skyrmion phases in magnetic systems \cite{rohart2013skyrmion,yang2023first}. However, its role in TTA remains unexplored. Notably, broadband ODMR has emerged as an early technique for quantitative measurement of the DMI strength in magnetic systems \cite{laplane2016}. In this study, we employ this technique to estimate the DMI interaction strength in a fully organic system. 

To introduce DMI we consider the extended Hamiltonian ${\hat H}(t) = {\hat H}_0(t) + \hat{H}_{DM}$ in which the DMI term $\hat{H}_{DM}$ is given by:
\begin{align}
  \hat{H}_{DM} = \mathbf{F} \cdot (  \mathbf{{\hat S}}_{a} \times \mathbf{{\hat S}}_{b} )
  \label{Htt}
\end{align}
The Hamiltonian $\hat{H}_{DM}$ is antisymmetric under the exchange of the two triplets and its only non-zero matrix elements are between the triplet and singlet/quintet manifolds. Its strength is characterized by a vector $\mathbf{F}$ depending on the microscopic positions of the two triplet excitons. A non-zero vector $\mathbf{F}$ requires the breaking of mirror symmetry, which can be achieved by dynamically inversion symmetry broken in TIPS-ADT, because the two triplet excitons in the triplet pair occupy different vibrational/excited states during their separation, even though the crystal structure belonging to the $P{\bar 1}$ centro-symmetric group. For this reason, we consider the $\mathbf{F}$ vector to be randomly oriented along the distance separating the two triplets and take an isotropic average over $\mathbf{F}$ vectors with a fixed magnitude. Since the DMI matrix elements are only non-zero for transitions between the triplet and joint singlet/quintet manifolds of the triplet pair, making their direct observation in magnetic resonance challenging. However we show that DMI can manifest indirectly through the induced triplet-triplet annihilation.

\begin{figure}
\includegraphics[clip=true,width=0.7\columnwidth]{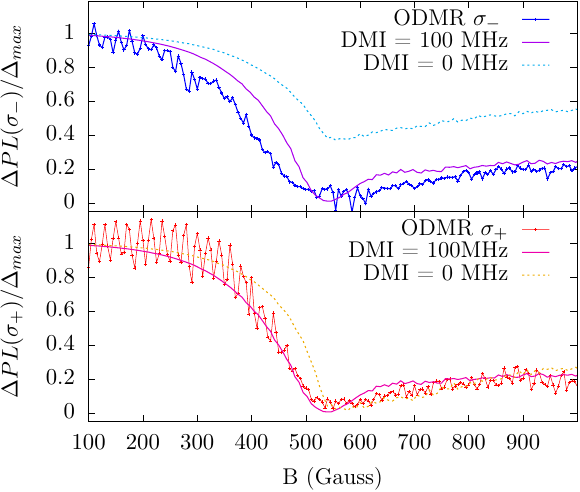}    
\caption{A comparison is made between the ODMR magnetic field dependence for $\sigma_{\pm}$ transitions, and the simulated population differences $P_0 - P_{\pm 1}$ within the TTA model, both with and without DMI interactions. The experimental data is normalized by the maximal ODMR amplitude (which is the same for both $\sigma_{\pm}$ transitions) while theoretical curves are normalized to $1$ at low field. A strong concordance is observed between the model and experimental results, particularly in capturing the weak asymmetry between $\sigma_{\pm}$ transitions in the presence of DMI.   }
\label{figtheoexp}
\end{figure}

Hyperfine interactions also create avoided level crossings with the triplet manifold. However, it brings a substantial asymmetry between $\sigma_{\pm}$ transitions. the hyperfine interaction can be modeled as two random magnetic fields acting separately on the two triplets corresponding to their respective nuclear spin environments. The amplitude of this random magnetic field is bounded by the magnetic resonance linewidth. If we take into account only hyperfine interactions without DMI, the simulations are shown in Fig.~\ref{figtheoexp}, where the ODMR amplitude from $\sigma_{-}$ transitions is much higher than the $\sigma_{+}$ transition after level crossing. For this reason, we focus on DMI-induced avoided level crossings in the following analysis.

\begin{figure}
\includegraphics[clip=true,width=0.7\columnwidth]{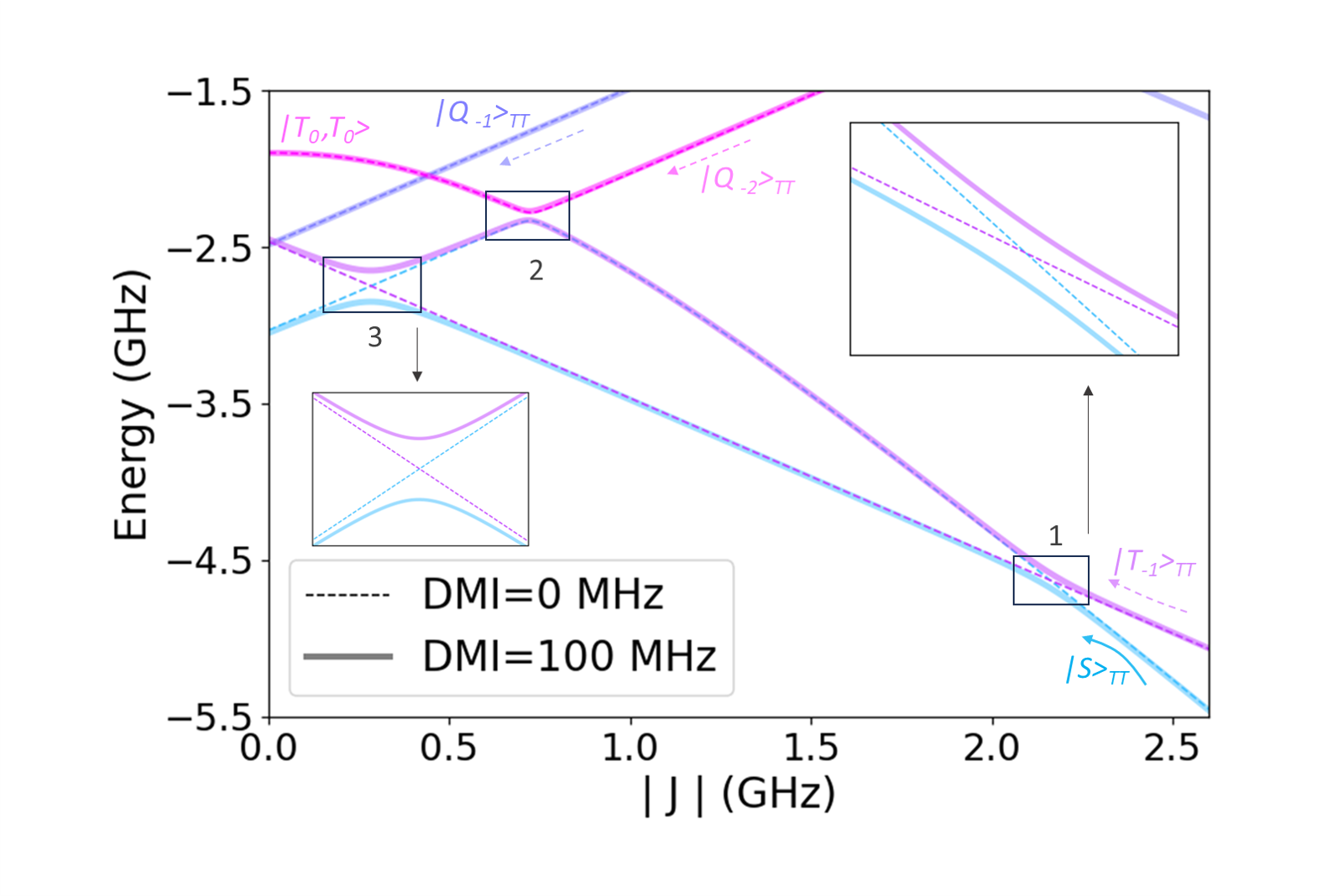}    
\caption{The energy diagram of the lowest energy states of a triplet pair as a function of the exchange energy $| J |$ is shown for a higher magnetic field $B = 1.4 D$ (low field case is shown on Fig.~\ref{figcrossinglowB}). In this regime, the triplet pair undergoes several possible level crossings as $|J|$ is lowered. These crossings, due to the Landau-Zener effect, lead to a decrease in the population transferred into the $|T_0, T_0\rangle$ state. As the $|J|$ decreases, the first level crossing (1) occurs with the triplet pair $|T_{-1}\rangle_{TT}$ state. A gap opens at this crossing, which is induced by a finite DMI strength. The population transferred into  $|T_{-1}\rangle_{TT}$  at this crossing is expected to undergo non-radiative TTA, thereby reducing the final population in the $|T_0,T_0\rangle$ state. The next level crossing (2) involves the quintet $|Q_{-2}\rangle_{TT}$, which is expected to lead to an increased population in $|T_{-1},T_{-1}\rangle$, although this is not observed experimently. This discrepancy is explained by the presence of a third level crossing (3), where $|Q_{-2}\rangle_{TT}$ intersects with  $|T_{-1}\rangle_{TT}$, and the gap for this crossing is also determined by the DMI, leading to TTA in this channel. 
}
\label{figcrossing}
\end{figure}

With a finite DMI, all the level crossings with the triplet manifold turn into anti-crossings. This is illustrated in Fig.~\ref{figcrossing}, where the eigen-spectrum as a function of exchange energy is represented by solid and dashed lines, corresponding to the cases with and without coupling, respectively, under the condition that $B = 700\;{\rm G} > D$.
The avoided crossings, impact the population transfer from the fission generated singlet $|S\rangle_{TT}$ during triplet dissociation. Following the triplet exciton pair from photogeneration to dissociation, Fig.~\ref{figcrossing} shows a series of avoided crossings as $|J(t)|$ drops to zero. The thick $|S\rangle_{TT}$ arrow shows the energy level followed by the initial singlet population while dashed arrows show initially empty states. The singlet population goes through a series of three avoided-level crossings. The first, labeled (1), occurs with the triplet manifold. With a finite DMI, (1) becomes an avoided crossing causing some population transfer into the triplet manifold which is then lost through TTA. The amount of population transfer will depend on whether the level crossing occurs in an adiabatic or diabatic manner, which is determined by the rate $dJ/dt$. The population remaining on the singlet branch will progress to the point labeled (2), which represents an avoided crossing, irrespective of the presence of DMI, as it involves the quintet manifold. The population transfer into the quintet state will cause an imbalance between the populations in $|T_{\pm1}\rangle$ giving an asymmetry in the amplitude of $\sigma_{\pm}$ which is not visible in the experiment. The finite DMI amplitude allows us to explain this lack of asymmetry. After point (2), the quintet line will cross the triplet line at point (3), which becomes an avoided crossing in the presence of DMI with a larger gap compared to points (1) and (2). This larger gap can lead to complete TTA of the population in $|Q_{-2}\rangle_{TT}$ suppressing the population imbalance between $|T_{\pm1}\rangle$ states. The remaining population for which both crossings at (1) and (2) are diabatic will transfer the population to the final state $|T_0, T_0\rangle$ leading to an equal ODMR signal for $\sigma_{\pm 1}$. As illustrated in Fig.~\ref{figcrossinglowB}.a, for low fields $B < 7D/9$, the singlet branch does not undergo any level crossing before reaching $|T_0, T_0\rangle$. In the range,  $7D/9< B < D$ as shown Fig.~\ref{figcrossinglowB}.b there are two crossings between  $|S\rangle_{TT}$ and $|T_{-1}\rangle_{TT}$ leading a drop in the final population of $|T_0, T_0\rangle$. At $B = D$,  the triplet manifold intersects the singlet branch precisely at $J = 0$. This is expected to result in a substantial increase in the TTA yield, as TTA can occur in a longer-lived quasi-separated state with low $J$. In contrast, for level crossings at finite $J$, TTA can only take place during a short-lived transient state after photogeneration. This phenomenon accounts for the vanishing ODMR signal at $B = D$.

By introducing a finite DMI, we qualitatively explain the key features of the ODMR spectrum shown in Fig.~\ref{odmrmap}. These features include: 1) The ODMR amplitude begins to decrease at 390 G, well below $B = D$, due to level crossings with the triplet manifold occurring when $B > 7D/9$. 2) ODMR vanishes at $B = D$ for completely separated pairs with $J=0$, where level crossing with the triplet manifold happens. 3) For $B > D$, the ODMR signal increases again, but to a smaller value than for $B < D$, without any population imbalance between $|T_{\pm1}\rangle$. We expect the excess population in $|Q_{-2}\rangle_{TT}$ to be completely lost due to TTA at the avoided level crossings 3, as seen in Fig.~\ref{figcrossing}. We now proceed with a quantitative analysis. For this purpose, we investigate numerically a theoretical model where we take into account TTA inside the triplet manifold adding a dissipative recombination term to the Schr\"odinger equation:
\begin{align}
  i \partial_t |\psi\rangle = {\hat H}(t) |\psi\rangle - \gamma_T {\hat P}_T |\psi\rangle
  \label{eq:sht}
\end{align}
where $|\psi\rangle$ is the wavefunction of the triplet pair and ${\hat P}_T$ is the projector onto the triplet manifold from the triplet pair Hilbert space. This model exclusively follows the dynamics of triplet exciton pairs, excluding the triplet excitons that remain after TTA. 
We note that the values of DMI and dipole-dipole interaction are important only during the short duration of level avoided-crossing. So we maintain these values as constants, even if they are expected to decay over time, analogous to the exchange energy $J(t)$. We initialize the wavefunction in the singlet $|\psi(t=0)\rangle = |S\rangle_{TT}$ with an initial exchange energy of $|J(t=0)| = 10\;{\rm GHz}$ being sufficiently far from any level crossings. We then assume that the exchange energy drops with a fixed rate $dJ/dt$ until $J(t) = 0$, after which we maintain $J = 0$ until equilibrium. We solve Eq.~(\ref{eq:sht}) numerically to determine the final state. At the end of the simulation, the final state $|\psi\rangle$ is projected onto the separated triplet states $|T_i, T_j\rangle$ ($i,j = 0, \pm 1$), from which we calculate the population $P_i$ of each triplet state $|T_i\rangle$. 

In Fig.~\ref{figtheoexp}, we compared the $\sigma_{\pm}$ amplitude from ODMR transitions in Fig.~\ref{odmrmap} with the populations obtained by solving Eq.~(\ref{eq:sht}) as a function of the magnetic field with and without DMI interaction. The amplitude of $\sigma_{+}$ transition is compared with the population difference $P_{0} - P_{+1}$, while $\sigma_{-}$ is compared with $P_0 - P_{-1}$. Both experimental data for $\sigma_{\pm}$ are normalized by their maximal amplitude $\Delta_{max}$. In addition to the smooth dependence, experimental curves show an oscillating part as a function of magnetic field. The latter comes from standing waves inside our microwave setup. To find the populations $P_{0, \pm 1}$, we average over randomly distributed rates $dJ/dt$ in the interval $20-60\; {\rm MHz/ns}$. This corresponds to a characteristic timescale for population transfer $D/(dJ/dt) \sim 35\;{\rm ns}$. For the dipole-dipole interaction, we assume a distance between the two triplets $d = 1.4\;{\rm nm}$ (second neighbor molecule distance) with a random orientation and a magnetic field aligned along the molecular $z$ axis (Fig.~\ref{odmrmap}) within $\pm 5^{\circ}$ accuracy. Hyperfine interactions were modeled by a Gaussian random magnetic field with variance $30\;{\rm MHz}$ comparable with the linewidth. Our simulations are in very good agreement with the experimental results for a DMI amplitude of $100\;{\rm MHz}$. The population in $P_0$ starts to drop at $B > 7 D/9$ due to the level crossing with the triplet manifold (see Figs.~\ref{figcrossinglowB} and \ref{figcrossing}). In the absence of DMI, hyperfine interactions provide the mixing required for TTA at this level crossing. It should be noted that without hyperfine interactions the decrease in ODMR amplitude starts much closer to $B = D$. However without DMI a strong asymmetry between $\sigma_{\pm}$ is obtained for $B > D$ which is not present in the experiment. The weak asymmetry in the experiment reaches its maximum around $B = D$ and is most likely due to the variation in spin lifetimes at around the level crossing between $|T_0\rangle$ and $|T_{-1}\rangle$ which is not taken into account in our TTA dynamical model.

In conclusion, we investigated the origin of spin-dependent fluorescence in single crystals of TIPS-ADT for a magnetic field aligned along the main fine structure direction. In this case, the magnetic resonance spectrum is well described by the linear Zeeman splitting of triplet excitons.  However, the magnetic field-dependent ODMR signal reveals the role of antisymmetric exchange DMI in triplet-separation explaining very accurately the evolution of the ODMR amplitudes with magnetic field. The DMI interaction induces level anticrossings between the triplet and both singlet or quintet manifolds as the triplet pair separates, leading to an additional non-radiative TTA channel. We showed a good agreement between experiments and a theoretical model, predicting the population dynamics of the triplet exciton spin eigenstates after triplet separation. This work leads to a better understanding of the role of spin-orbit couplings in the interaction between triplet excitons, which appear in many photo-physical processes. The effect of DMI is only important at avoided level crossings with the triplet manifold of the exciton pair, and manifests rather indirectly through the induced change in the spin populations and enhanced TTA. It is thus possible that DMI plays a more important role in the photo-physics of triplet pairs that so far acknowledged.

{\bf Acknowledgements:} This project was supported by funding from ANR-20-CE92-0041 (MARS) and IDF-DIM SIRTEQ, and the European Research Council (ERC) under the European Union’s Horizon 2020 research and innovation program (grant Ballistop agreement no. 833350). V. Derkach acknowledges kind hospitality from CNRS Gif-sur-Yvette.

\bibliography{dmiodmr.bib}

\end{document}